\documentclass[prb,twocolumn]{revtex4}%

\usepackage{graphicx}
\usepackage{amsmath}
\usepackage{ulem}
\usepackage{color}

\newcommand{\be}{\begin{equation}}
\newcommand{\ee}{\end{equation}}
\newcommand{\bea}{\begin{eqnarray*}}
\newcommand{\eea}{\end{eqnarray*}}
\newcommand{\bean}{\begin{eqnarray}}
\newcommand{\eean}{\end{eqnarray}}

\begin{document}

\draft
\title{\bf
Impact of Valley Degeneracy on Thermoelectric Properties of Zigzag
Graphene Nanoribbons with Staggered Sublattice Potentials and
Transverse Electric Fields}

\author{David M T Kuo}

\address{Department of Electrical Engineering and Department of Physics, National Central
University, Chungli, 32001 Taiwan, Republic of China}

\date{\today}

\begin{abstract}
This study investigates the band inversion of flat bands in zigzag
graphene nanoribbons (ZGNRs) using a tight-binding model. The band
inversion results from symmetry breaking in the transverse
direction, achievable through deposition on specific substrates
such as separated silicon carbide or hexagonal boron nitride
sheets. Upon band inversion, ZGNRs exhibit electronic structures
characterized by valley degeneracy and band gap properties, which
can be modulated by transverse electric fields. To explore the
impact of this level degeneracy on thermoelectric properties, we
employ Green's function techniques to calculate thermoelectric
quantities in ZGNR segments with staggered sublattice potentials
and transverse electric fields. Two carrier transport scenarios
are considered: the chemical potential is positioned above and
below the highest occupied molecular orbital. We analyze
thermionic-assisted transport (TAT) and direct ballistic transport
(DBT). Level degeneracy enhances the electric power factors of
ZGNRs by increasing electrical conductance, while the Seebeck
coefficient remains robust in the TAT scenario. Conversely, in
DBT, the enhancement of the power factor primarily stems from
improvements in the Seebeck coefficient at elevated temperatures.
\end{abstract}

\maketitle

\section{Introduction}
Thermoelectric heat engines (TEHEs) function as electrical
generators when exposed to temperature differentials and can
alternatively operate as coolers when electric currents are
applied [\onlinecite{MahanGD}--\onlinecite{Minnich}]. This dual
capability makes TEHEs a promising solution for addressing
environmental concerns related to CO$_2$ emissions. High-quality
thermoelectric materials exhibit electron crystal-like properties
and act as phonon insulators
[\onlinecite{MahanGD}-\onlinecite{MahanG}], characterized by long
electron mean free paths and short phonon mean free paths.
Consequently, extensive research has focused on exploring the
thermoelectric properties of low-dimensional systems
[\onlinecite{Siivola}--\onlinecite{LinYM}]. Materials composed of
low-dimensional structures such as quantum wells
[\onlinecite{Siivola}], quantum wires
[\onlinecite{Boukai}--\onlinecite{TianY}], and quantum dots
[\onlinecite{Harman}] demonstrate significant enhancements in
thermoelectric properties. These structures effectively increase
phonon scattering rates at interfaces, thereby reducing phonon
thermal conductance, while maintaining electron thermoelectric
properties comparable to bulk materials
[\onlinecite{Siivola}--\onlinecite{LinYM}].

The efficiency of thermoelectric materials is quantified by the
figure of merit $ZT=S^2G_eT/\kappa$, where $S$, $G_e$ and $T$
denote the Seebeck coefficient, electrical conductance, and
equilibrium temperature, respectively. Thermal conductance
$\kappa$ comprises contributions from electron thermal conductance
$\kappa_e$ and phonon thermal conductance $\kappa_{ph}$. Initial
theoretical work by Hicks and Dresselhaus, using the effective
mass model, suggested that quasi-one-dimensional systems with
diameters less than one nanometer could achieve $ZT = 14$
[\onlinecite{Hicks}], sparking extensive research into the
thermoelectric properties of various material quantum wires and
those exhibiting one-dimensional topological states using
first-principles methods
[\onlinecite{OuyangYJ}--\onlinecite{ChangPH}].

The landmark discovery of graphene in 2004 by Novoselov and Geim
[\onlinecite{Novoselovks}] prompted investigations into
quasi-one-dimensional electronics, optoelectronics, and
thermoelectric devices utilizing nanowires formed from
two-dimensional materials [\onlinecite{WangHM},
\onlinecite{Chunhua}]. Graphene nanoribbon (GNR)-based devices
include field-effect transistors [\onlinecite{PassiV}],
spintronics [\onlinecite{CaoL}], gas sensors [\onlinecite{LiuGG}],
nanoscale rotators [\onlinecite{LiXB}], and electrical diodes
[\onlinecite{DingWC}]. GNRs have also garnered attention in
thermoelectric applications due to their low thermal conductance,
as reported in theoretical calculations [\onlinecite{OuyangYJ}],
which is attributed to their atomic-scale cross-section.
Edge-disordered zigzag GNRs (ZGNRs) [\onlinecite{SevincliH}],
semiconducting armchair GNRs (AGNRs) [\onlinecite{MazzamutoF}],
and GNRs with topological states [\onlinecite{XuY}] exhibit
significantly enhanced $ZT$ values compared to gapless graphenes
[\onlinecite{ChangPH}]. These GNRs achieve $ZT$ values greater
than one, although their power factors are relatively small. TE
devices require $ZT$ values exceeding three, along with optimized
electrical power outputs [\onlinecite{Whitney}]. Enhancing the
power factor values through increased level degeneracy has been a
focal point of numerous studies
[\onlinecite{Kuo1}--\onlinecite{Ranganayakulu}]. Materials like
$GeTe$ exhibit very high $ZT$ at elevated temperatures, where
valley degeneracy may play a crucial role
[\onlinecite{Toriyama}--\onlinecite{Ranganayakulu}].

While bottom-up techniques provide precise atomic control over
various GNRs, the existence of semiconducting GNRs with valley
degeneracy has not yet been reported
[\onlinecite{Cai}--\onlinecite{JeilJ}]. To investigate the impact
of valley degeneracy on the thermoelectric properties of GNRs, we
propose utilizing ZGNRs with staggered sublattice potentials.
These potentials can be implemented by coupling the ZGNRs to
substrates such as silicon carbide or hexagonal boron-nitride
[\onlinecite{ZhaoJ}-\onlinecite{ZhouSY}].Most recently, an
experimental report indicated a band gap of $E_g = 0.6$~eV for
graphene, attributed to the effect of silicon carbide substrates
[\onlinecite{ZhaoJ}]. This impressive $E_g$ value is significantly
larger than theoretical predictions [\onlinecite{Giovannetti},
\onlinecite{ZhouSY}]. In Fig. 1, we consider three scenarios of
zigzag graphene nanoribbons (ZGNRs) with staggered sublattice
potentials, denoted as (a) w-AB/n-C/w-AB, (b) w-AB/n-C/w-BA, and
(c) w-BA/n-C/w-AB. Here, the integers w and n represent the widths
of ZGNRs with and without sublattice potential, respectively.
Sublattices A and B are subjected to the staggered potentials. One
of the crucial findings is that the w-AB/4-C/w-AB structures
exhibit semiconducting behavior and valley degeneracy in two
valleys within the conduction and valence subbands. This valley
degeneracy arises from band inversion of the flat bands induced by
inversion symmetry breaking in the transverse direction,
facilitated by the underlying substrates. The primary objective of
this study is to investigate how valley degeneracy affects the
thermoelectric properties of ZGNRs supported on these substrates.

\begin{figure}[h]
\centering
\includegraphics[trim=1.cm 0cm 1.cm 0cm,clip,angle=0,scale=0.33]{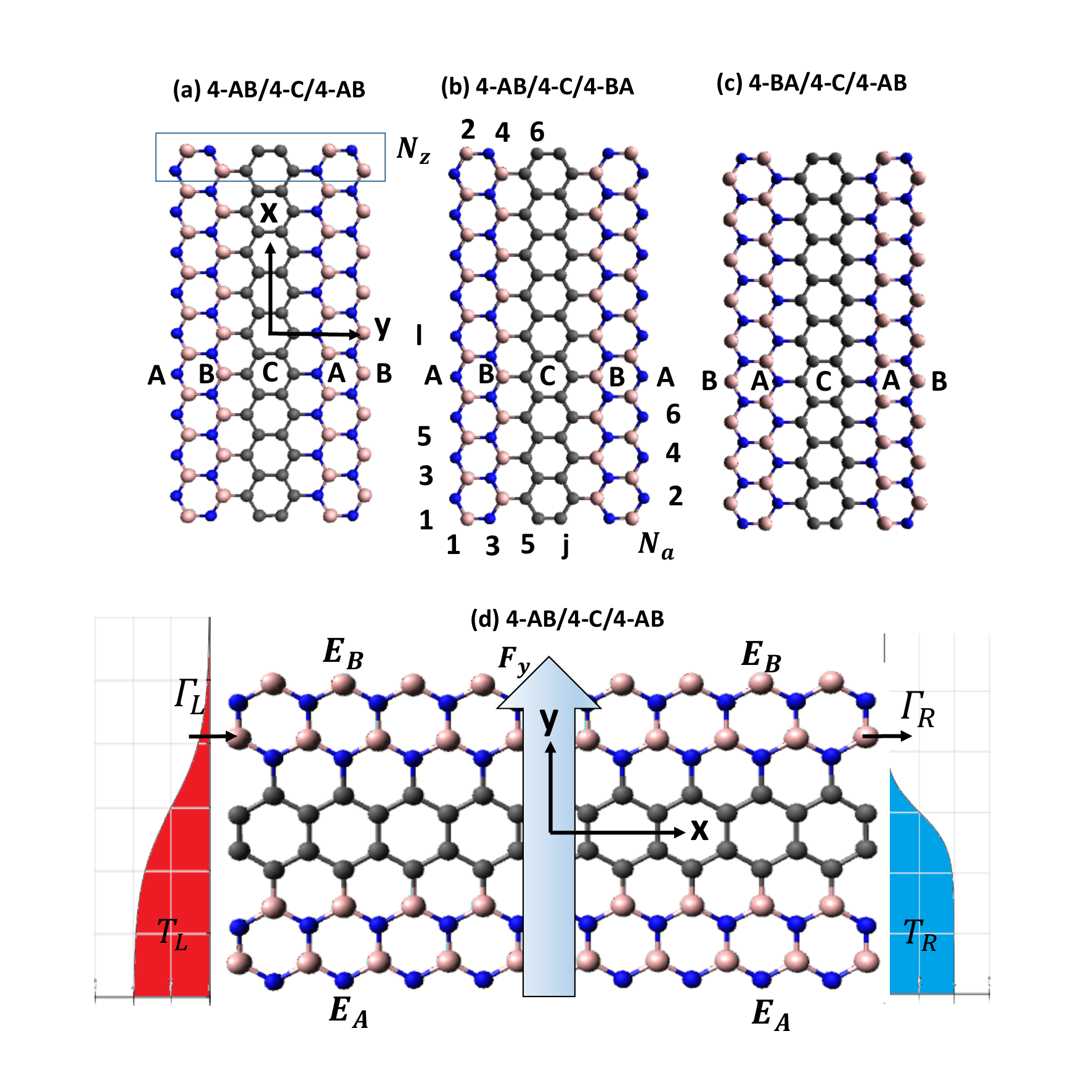}
\caption{ Schematic diagram illustrating three types of ZGNRs with
staggered sublattice potentials: (a) 4-AB/4-C/4-AB, (b)
4-AB/4-C/4-BA, and (c) 4-BA/4-C/4-AB. In (a), ZGNR without
inversion symmetry in the y direction. The square bracket
indicates the unit cell of 4-AB/4-C/4-AB. In (b) and (c), ZGNRs
remain the inversion symmetry in the y direction.  Atoms marked in
blue and white indicate that these carbon atoms are subjected to
staggered potentials $E_A = \delta$ and $E_B = -\delta$.
Additionally, $E_C = 0$ denotes the $p_z$ orbital of the carbon
atom. Diagram (d) depicts the armchair edges of the 4-AB/4-C/4-AB
segment in contact with electrodes. Symbols $\Gamma_{L}$
($\Gamma_R$) represent the electron tunneling rate between the
left (right) electrode and the leftmost (rightmost) atoms at the
armchair edges, and $T_{L}$ ($T_{R}$) denotes the equilibrium
temperature of the left (right) electrode. $F_y$ denotes a uniform
transverse electric field along the armchair direction.}
\end{figure}

\section{Calculation Methodology}
To explore the thermoelectric properties of ZGNRs with sublattice
potentials connected to the electrodes, we utilize a combination
of the tight-binding model and the Green's function technique. The
system Hamiltonian consists of two components: $H = H_0 +
H_{ZGNR}$. Here, $H_0$ signifies the Hamiltonian of the
electrodes, encompassing the interaction between the electrodes
and the ZGNR. Meanwhile, $H_{ZGNR}$ represents the Hamiltonian for
the ZGNRs and can be expressed as follows:

\begin{small}
\begin{eqnarray}
H_{ZGNR}&= &\sum_{\ell,j} E_{\ell,j} d^{\dagger}_{\ell,j}d_{\ell,j}\\
\nonumber&-& \sum_{\ell,j}\sum_{\ell',j'} t_{(\ell,j),(\ell', j')}
d^{\dagger}_{\ell,j} d_{\ell',j'} + h.c,
\end{eqnarray}
\end{small}

Here, $E_{\ell,j}$ represents the on-site energy of the orbital in
the ${\ell}$-th row and $j$-th column. The operators
$d^{\dagger}_{\ell,j}$ and $d_{\ell,j}$ create and annihilate an
electron at the atom site denoted by ($\ell$,$j$). The parameter
$t_{(\ell,j),(\ell', j')}$ characterizes the electron hopping
energy from site ($\ell'$,$j'$) to site ($\ell$,$j$). We set
$t_{(\ell,j),(\ell',j')} = t_{pp\pi} = 2.7$ eV for the
nearest-neighbor hopping strength. Because we have considered
ZGNRs on the hexagonal boron nitride substrates, staggered
sublattice potentials are $E_{A} = \delta$ and $E_{B} = -\delta$,
which are relative to $p_z$ orbital of carbon atom $E_C =
0$[\onlinecite{Giovannetti},\onlinecite{QiaoZH}]. Note that if one
stack ZGNRs on other two dimensional material substrates and then
spin orbital coupling may not be ignored[\onlinecite{LuCP}]. When
a uniform transverse electric field $E_F$ is introduced into the
system, electrical potential $ U = e F_y y$ affects each site of
the ZGNRs. Here, $F_y = V_y/L_a$, where $V_y$ is the applied bias
and $L_a$ is the width of the ZGNRs [\onlinecite{Pizzochero}].

As mentioned in the introduction, the efficiency of thermoelectric
materials is determined by the dimensionless figure of merit $ZT =
S^2G_eT/(\kappa_e+\kappa_{ph})$[\onlinecite{LinYM},\onlinecite{Hicks}]
where electrical conductance $G_e$, Seebeck coefficient$S$, and
electron thermal conductance $\kappa_e$ can be computed using
$G_e=e^2{\cal L}_{0}$, $S=-{\cal L}_{1}/(eT{\cal L}_{0})$ and
$\kappa_e= \frac{1}{T}({\cal L}_2-\frac{{\cal L}^2_1}{{\cal
L}_0})$ with ${\cal L}_n$ ($n=0,1,2$) defined as

\begin{equation}
{\cal L}_n=\frac{2}{h}\int d\varepsilon~ {\cal
T}_{LR}(\varepsilon)(\varepsilon-\mu)^n\frac{\partial
f(\varepsilon)}{\partial \mu}.
\end{equation}

Here, $f(\varepsilon) = 1/(1+\exp((\varepsilon-\mu)/k_BT))$
represents the Fermi distribution function of electrodes at
equilibrium chemical potential $\mu$. The constants $e$, $h$,
$k_B$, and $T$ denote the electron charge, Planck's constant,
Boltzmann's constant, and the equilibrium temperature of the
electrodes, respectively. ${\cal T}_{LR}(\varepsilon)$ signifies
the transmission coefficient of a ZGNR connected to electrodes,
and it can be calculated using the formula ${\cal
T}_{LR}(\varepsilon) =
4Tr[\Gamma_{L}(\varepsilon)G^{r}(\varepsilon)\Gamma_{R}(\varepsilon)G^{a}(\varepsilon)]$
[\onlinecite{Kuo3},\onlinecite{SunQF}], where
$\Gamma_{L}(\varepsilon)$ and $\Gamma_{R}(\varepsilon)$ denote the
tunneling rate (in energy units) at the left and right leads,
respectively, and ${G}^{r}(\varepsilon)$ and
${G}^{a}(\varepsilon)$ are the retarded and advanced Green's
functions of the GNRs, respectively. The tunneling rates
($\Gamma_{L(R)}(\varepsilon)= Im\sum^r_{L(R)}(\varepsilon)$) are
determined by the imaginary part of the self-energy originating
from the coupling between the left (right) electrode and its
adjacent GNR atoms. In terms of tight-binding orbitals,
$\Gamma_{\alpha}(\varepsilon)$ and Green's functions are matrices.
For simplicity, $\Gamma_{\alpha}(\varepsilon)$ for interface atoms
possesses diagonal entries with a common value of $\Gamma_t$
[\onlinecite{Kuo3}]. Contact properties between 2D nanostructures
and metallic electrodes typically induce Schottky barriers
[\onlinecite{Areshkin}--\onlinecite{Matsuda}], posing challenges
in achieving tunable Fermi energy due to Fermi energy pinning
[\onlinecite{ChenRS}]. Despite numerous theoretical studies aiming
to elucidate this crucial behavior from first principles, the
inherent limitations result in only qualitative insights regarding
$\Gamma_t$ arising from the contact junction
[\onlinecite{LeeG},\onlinecite{Matsuda}]. Therefore, the
tight-binding model serves as an effective approach for
calculating the electronic structures and charge transport of GNRs
[\onlinecite{Kuo1}-\onlinecite{Kuo3},\onlinecite{JeilJ},\onlinecite{Pizzochero}].
In this work, we developed Fortran computational code to evaluate
the electronic structures, transmission coefficients, and
thermoelectric quantities of ZGNRs with staggered sublattice
potentials and transverse electric fields.

\section{Results and discussion}
\subsection{Electronic Structures of ZGNRs with Staggered Sublattice Potentials}

To depict the electronic properties of ZGNRs with staggered
sublattice potentials, as shown in Fig. 1(a)-1(c), we calculated
the electronic structures of three configurations: 4-AB/4-C/4-AB,
4-BA/4-C/4-AB, and 4-AB/4-C/4-BA, based on the Hamiltonian in Eq.
(1). These electronic structures are presented in Fig. 2(a), 2(b),
and 2(c), respectively. Among ZGNRs with sublattice potentials,
only the w-AB/4-C/w-AB structure exhibits a semiconducting phase.
Notably, the w-AB/4-C/w-AB structure also displays valley
degeneracy in the frontier conduction and valence subbands, albeit
with shallow valley depths. In contrast, both the 4-BA/4-C/4-AB
and 4-AB/4-C/4-BA configurations exhibit metallic phases,
characterized by asymmetrical electronic structures distinct from
those of ZGNRs without staggered sublattice potentials, where
symmetrical flat bands are observed near the charge neutral point
(CNP) [\onlinecite{Nakada},\onlinecite{Wakabayashi}].

Figures 2(d), 2(e), and 2(f) illustrate the electronic structures
of w-AB/4-C/w-AB for three different widths, where w $= 8 , 12 $
and $16$. The valley structure of the frontier conduction and
valence subbands is sensitive to the widths of ZGNRs coupled to
the substrates. Based on the findings of Fig. 2, a semiconducting
phase with valley degeneracy emerges from the w-AB/4-C/w-AB
structure due to broken inversion symmetry in the transverse
direction. However, this valley structure is suppressed by the
coupling between the frontier and secondary conduction (valence)
subbands. The mechanism of valley degeneracy in topological
thermoelectric materials remains a topic of current interest
[\onlinecite{ToriyamaMY}-\onlinecite{ZhuZY}].

\begin{figure}[h]
\centering
\includegraphics[angle=0,scale=0.3]{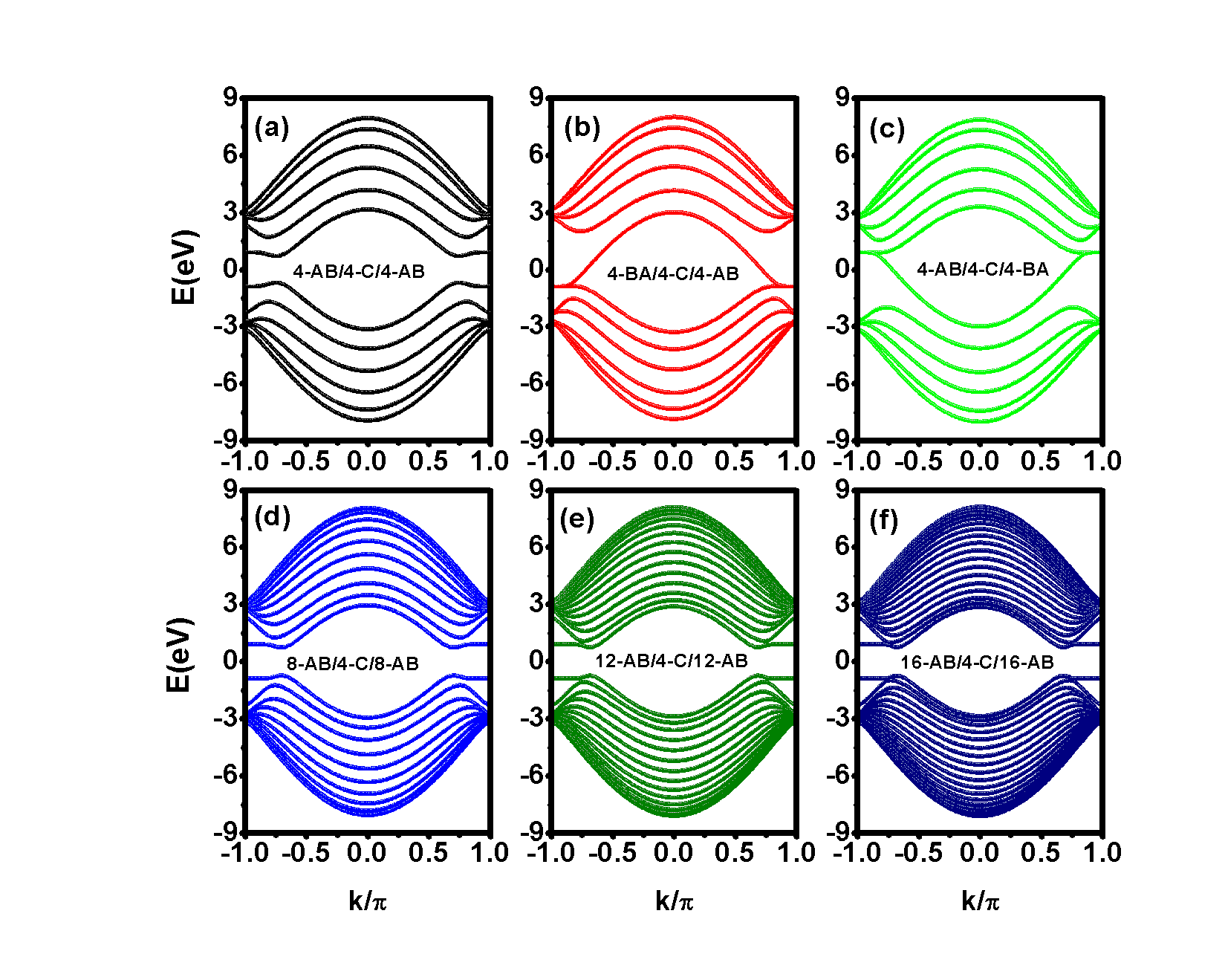}
\caption{Electronic structures of ZGNRs with various
configurations: (a) 4-AB/4-C/4-AB, (b) 4-BA/4-C/4-AB, (c)
4-AB/4-C/4-BA, (d) 8-AB/4-C/8-AB, (e) 12-AB/4-C/12-AB, and (f)
16-AB/4-C/16-AB. Here, we have adopted $E_A = -E_B = \delta =
0.9$~eV.}
\end{figure}

When transverse electric fields $F_y$ are applied to ZGNRs, the
electric potential $ U = e V_y y/L_a$ affects each site of the
ZGNRs [\onlinecite{Pizzochero}]. This indicates that $E_A$, $E_B$
and $E_C$ depend on the strength and direction of $V_y$. We
investigate the electronic structures of the 4-AB/4-C/4-AB
structure under transverse electric fields, as depicted in Fig. 3.
Increasing $V_y$ results in a reduction of the band gap in the
4-AB/4-C/4-AB structure. At $V_y = 0.9$ V
 , the electronic structures of the frontier conduction and valence subbands exhibit flat bands. Conversely, negative values of
$V_y$ reveal more pronounced valley structures in Fig. 3(d), 3(e),
and 3(f). This negative $ V_y $ enhances the inversion symmetry
breaking of ZGNRs [\onlinecite{ZhuZY}]. To illustrate reduction
and enhancement of the band gap, let's consider the sublattice
sites at $j = 1$ with energy level $E_A$ and $j = N_a$ with energy
level $E_B$. For $V_y
> 0$, the sublattice potentials are modified to $\delta_A = \delta-eV_y/2$ and
$\delta_B = -\delta+eV_y/2 = - \delta_A$. For $V_y < 0$, the
potentials $E_A=\delta$ and $E_B=-\delta$ are adjusted to
$\delta_A=\delta+eV_y/2$ and $\delta_B=-\delta-eV_y/2=-\delta_A$.
The enhancement and reduction of the sublattice potentials
directly influence the corresponding changes in the band gaps. The
directionality of the electric field influences the valley
structure of 4-AB/4-C/4-AB ZGNRs with asymmetrical staggered
sublattice potentials. Such an electric field-dependent valley
structure could potentially find applications in valleytronics
[\onlinecite{Schaibley}-\onlinecite{Motohiko}].

\begin{figure}[h]
\centering
\includegraphics[angle=0,scale=0.3]{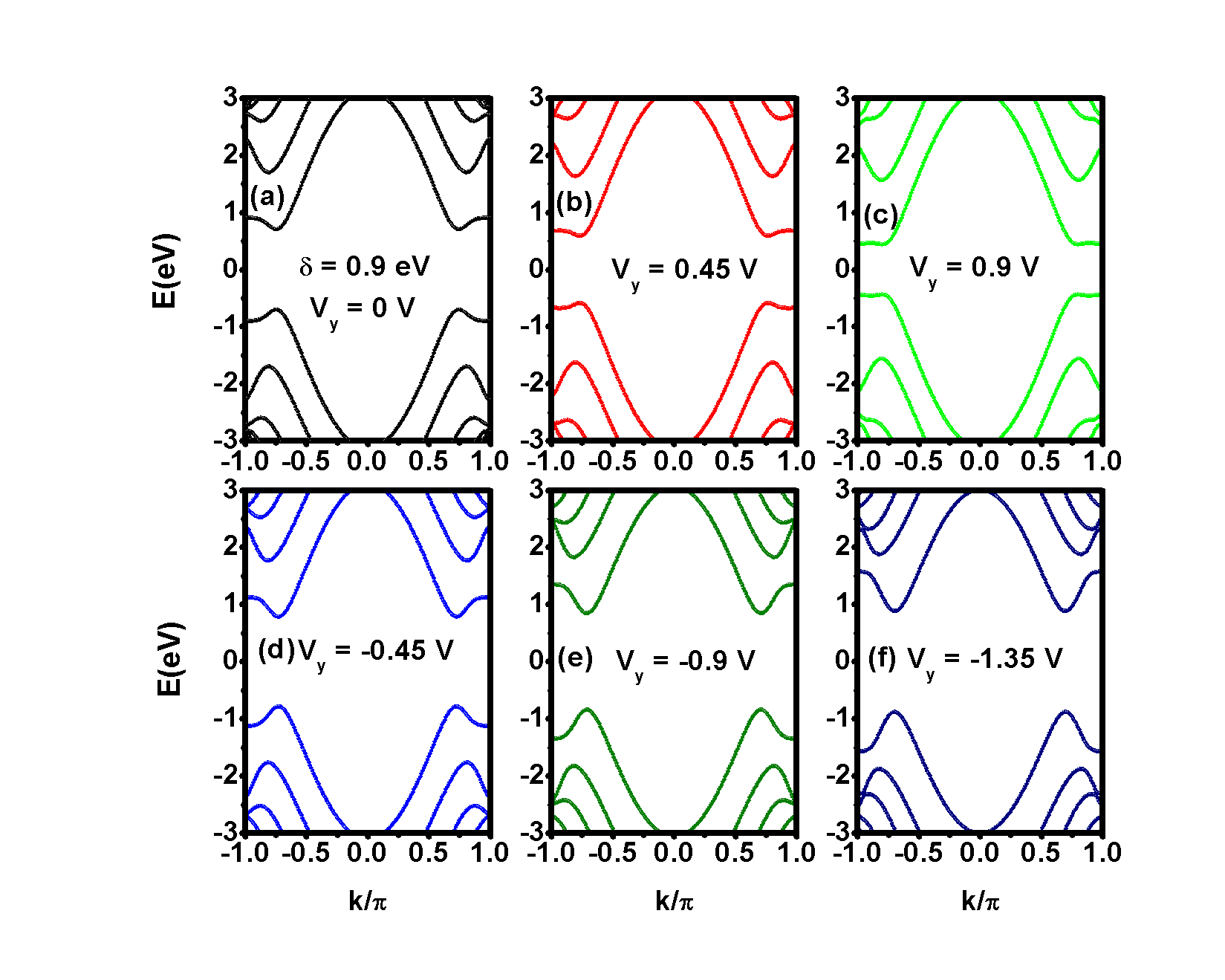}
\caption{Electronic structures of 4-AB/4-C/4-AB ZGNRs with
staggered sublattice potential $\delta = 0.9$~eV for various
transverse applied voltages: (a) $V_y = 0$~V, (b) $V_y = 0.45$~V,
(c) $V_y = 0.9$~V, (d) $V_y = -0.45$~V, (e) $V_y = -0.9 $~V and
(f) $V_y = -1.35$~V.}
\end{figure}

\subsection{Transmission Coefficient of 4-AB/4-C/4-AB under Transverse Electric Fields}
Synthesizing ZGNRs longer than 20 nm using bottom-up techniques
remains challenging [\onlinecite{LlinasJP},
\onlinecite{JacobsePH}]. Therefore, investigating the size effect
of ZGNR segments on valley degeneracy is crucial. Here, we present
the energy levels of ZGNR segments with $N_z = 81 $ ( $L_z
 = 9.84 $nm) as functions of the sublattice potential magnitude $\delta$
under various transverse electric fields in Fig. 4. When $\delta =
0$ and $V_y = 0$, metallic ZGNRs exhibit discrete energy levels
due to their finite sizes, as seen in Fig. 4(a). The energy levels
$E_{HO}$ (highest occupied molecular orbital, HOMO) and $E_{LU}$
(lowest unoccupied molecular orbital, LUMO) are very close to the
charge neutral point (CNP) due to small overlap between the wave
functions of zigzag edge states. As $\delta $ increases, $E_{HO}$
and $E_{LU}$ separate. Moreover, the number of level degeneracies,
indicated by a wider range of levels with stronger color
intensity, increases for $\delta > 0.45 $~eV. These results in
Fig. 4(a) can be understood in the context of the electronic
structures of 4-AB/4-C/4-AB ZGNRs shown in Appendix A (Fig. A.1).
The extent of level degeneracy in 4-AB/4-C/4-AB ZGNR segments
corresponds to the depth of valley structures in infinite
4-AB/4-C/4-AB ZGNRs. Upon applying transverse electric fields, the
patterns of energy levels with respect to $\delta$ change
dramatically, as illustrated in Figs. 4(b), 4(c), and 4(d).
Because the energy levels $E_{HO}$ and $E_{LU}$ exhibit very high
probability densities at the zigzag edge sites, specifically at $j
= 1$ and $j = N_a$ (see Fig. 5), $V_y$-dependent $E_{HO}$ and
$E_{LU}$ can be readily understood in terms of $\delta_A$ and
$\delta_B$ as illustrated in Fig. 3. For example, $E_{LU}$ and
$E_{HO}$ very close to CNP at $\delta = 0.45$~eV and $V_y =
0.9$~V, as shown in Fig. 4(b). The range of energy level
degeneracy with respect to transverse electric fields aligns with
the electronic structures shown in Fig. 3. Notably, while ZGNRs
without staggered sublattice potential have demonstrated valley
structures under strong transverse electric fields, the extent of
level degeneracy is limited, and the electric
field-direction-dependent valley structure disappears
[\onlinecite{LeeKW}].

\begin{figure}[h]
\centering
\includegraphics[angle=0,scale=0.3]{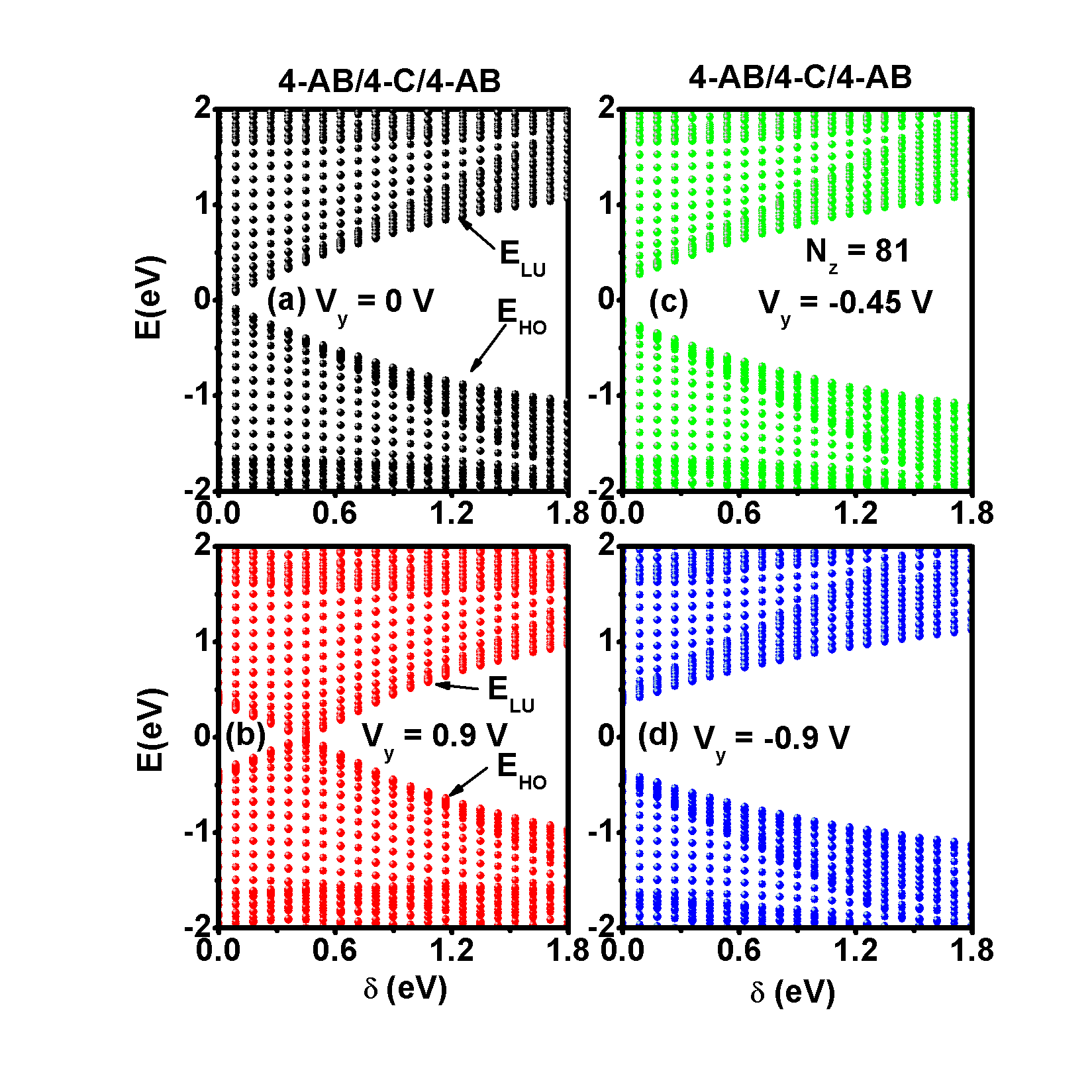}
\caption{Energy levels of 4-AB/4-C/4-AB ZGNR segments as functions
of $\delta$ for various transverse applied voltages: (a) $V_y =
0$~V, (b) $V_y = 0.9$~V, (c) $V_y = -0.45$~V, and (d) $V_y =
-0.9$~V. The ZGNR segment with length $L_z$ is characterized by
$L_z = 9.84$~nm ($N_z = 81$).}
\end{figure}

Since the transmission coefficient spectra of nanostructures are
influenced by the wave-functions of their energy levels
[\onlinecite{Kuo3}], understanding the wave functions of each
energy level for 4-AB/4-C/4-AB ZGNR segments with various $\delta$
values is crucial. In Fig. 5(a), the probability densities
$|\Psi(\ell,j)(E = 0)|^2$ demonstrate localized and symmetrical
zigzag edge states in the $ y$ direction for $\delta = 0$~eV and
$V_y = 0$. Figures 5(b) and 5(c) show the probability densities of
$E_{HO} = -0.253$~eV and $E_{LU} = 0.253$~eV, respectively,
depicting the wave functions of the HOMO and the LUMO at $\delta =
0.27$~eV. Analysis of their probability density distributions
reveals that $E_{HO(LU)} = \mp 0.253$~eV are primarily determined
by $E_B$ and $E_A$, respectively. The maximum probability
densities $|\Psi(\ell,j)(E_{HO})|^2$ and
$|\Psi(\ell,j)(E_{LU})|^2$ are concentrated at $ j = 12 $ and $ j
= 1$, respectively, with $|\Psi(\ell,j)(E_{HO})|^2$ being
vanishingly small for $j = 7 , 9 $ and $11 $, and
$|\Psi(\ell,j)(E_{LU})|^2$ for $ j = 2, 4$ and $6$. In Fig. 5(d),
5(e) and 5(f) at $\delta = 0.9$~eV, we present the probability
density distributions of $E_1 = -0.704$~eV, $E_2 = -0.7053$~eV and
$E_3 = -0.7284$~eV, which are below the CNP. These states exhibit
delocalized and asymmetrical characteristics in the $ y$
direction, indicative of band inversion in flat bands for ZGNRs.
The probability densities $|\Psi(\ell = 1,j)(E_i)|^2$ and
$|\Psi(\ell = 81,j)(E_i)|^2$ significantly influence the
transmission coefficient spectra of 4-AB/4-C/4-AB ZGNR segments
with armchair edges coupled to the electrodes shown in Fig. 1(d).

\begin{figure}[h]
\centering
\includegraphics[angle=0,scale=0.35]{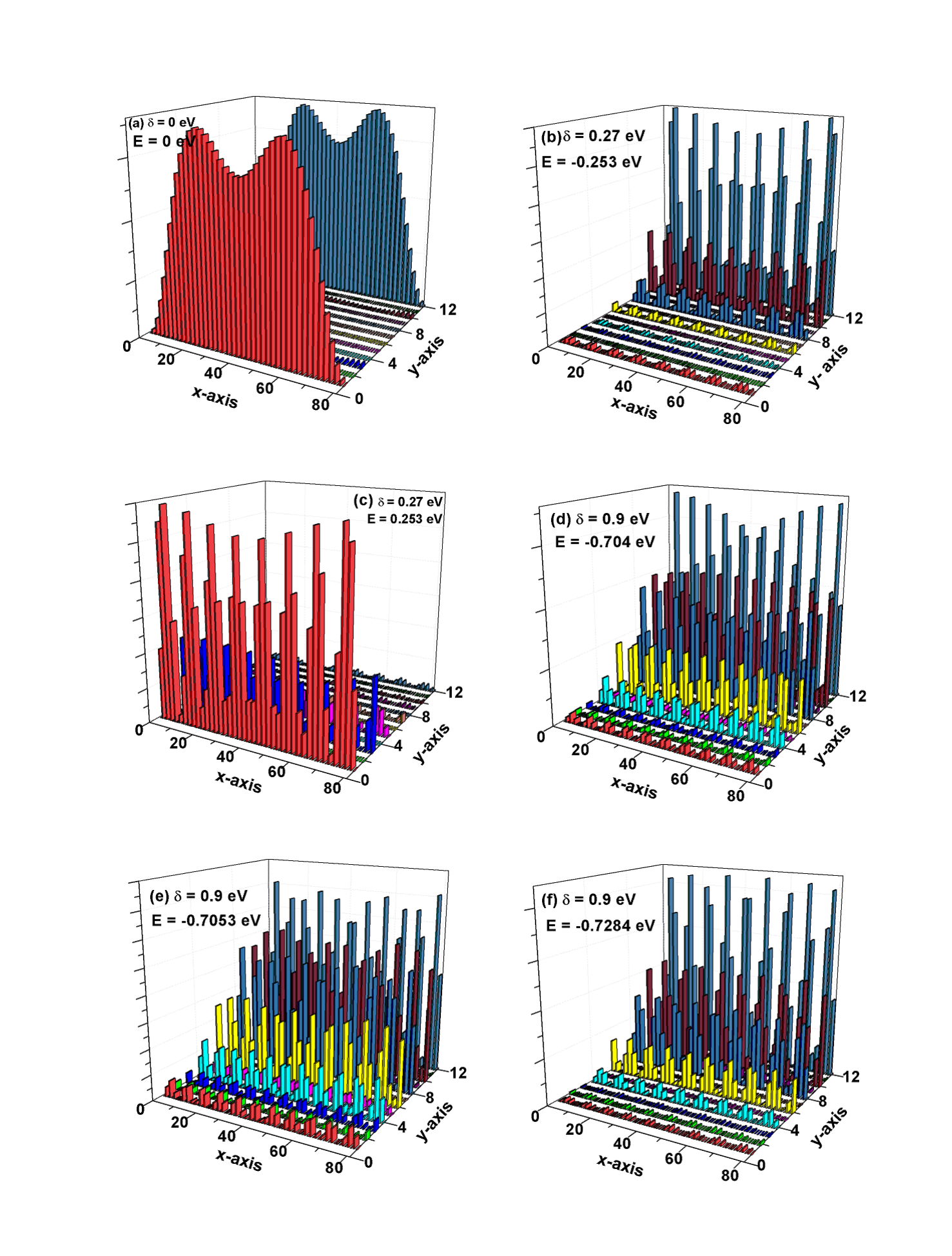}
\caption{Probability density of 4-AB/4-C/4-AB ZGNRs with different
staggered sublattice potentials in the absence of $V_y$: (a)
$\delta = 0$ and $E = 0$, (b) $\delta = 0.27$~eV and $E_{HO} =
-0.253$~eV, (c) $\delta = 0.27$~eV and $E_{LU} = 0.253$~eV, (d)
$\delta = 0.9$~eV and $E_1 = -0.704$~eV,(e) $\delta = 0.9$~eV and
$E_2 = -0.7053$~eV, and (f) $\delta = 0.9$~eV and $E_3 =
-0.7284$~eV.}
\end{figure}

According to Eq. (2), the thermoelectric quantities of ZGNRs,
including electrical conductance, Seebeck coefficient, and
electron thermal conductance, are determined by the area and shape
of the transmission coefficient (${\cal T}_{LR}(\varepsilon)$)
curve. Therefore, it is necessary to clarify the effects of
contact properties on ${\cal T}_{LR}(\varepsilon)$. In Fig. 6, we
present the calculated ${\cal T}_{LR}(\varepsilon)$ curves of
4-AB/4-C/4-AB ZGNR segments for various $\Gamma_t$ values, which
depend on the contact geometries and the electrode materials. For
simplicity of analysis, $\varepsilon \le 0$ is considered
throughout this article. Due to electron-hole symmetry, the curves
of transmission coefficients for $\varepsilon \ge 0$ are readily
obtained (or see Fig. B.1). As seen in Fig. 6(a), the ${\cal
T}_{LR}(\varepsilon)$ curve shows many discrete peaks under a weak
coupling strength of $\Gamma_t = 0.27$~eV. With increasing
$\Gamma_t$ values, the area of ${\cal T}_{LR}(\varepsilon)$ curve
increases, and the shapes of these areas also change. For
instance, in Fig. 6(b), the area of ${\cal T}_{LR}(\varepsilon)$
curve shows a right-triangle shape with a steep change with
respect to $\varepsilon$  on the side toward the central gap. In
Figs. 6(c) and 6(d), the areas of ${\cal T}_{LR}(\varepsilon)$
curves show rectangular shapes. For $\Gamma_t = 2.7$~eV, we
observe an arch shape in Fig. 6(f). The level degeneracy is
characterized by  ${\cal T}_{LR}(\varepsilon)$ with values near
two. For $\Gamma_t = 2.7$~eV, the electrode materials can be
regarded as graphenes since $\Gamma_t = t_{pp\pi}$. Because the
spectra of the transmission coefficient ${\cal
T}_{LR}(\varepsilon)$ involve the Green's functions, the local
density of states determined by the imaginary part of
$G^{r(a)}(\varepsilon)$ directly reveals the $\Gamma_t$-dependent
probability density of each energy level in Fig. 6, which
determines the shape of the transmission coefficient curve. We
refer readers to our previous work[\onlinecite{Kuo3}].

\begin{figure}[h]
\centering
\includegraphics[angle=0,scale=0.3]{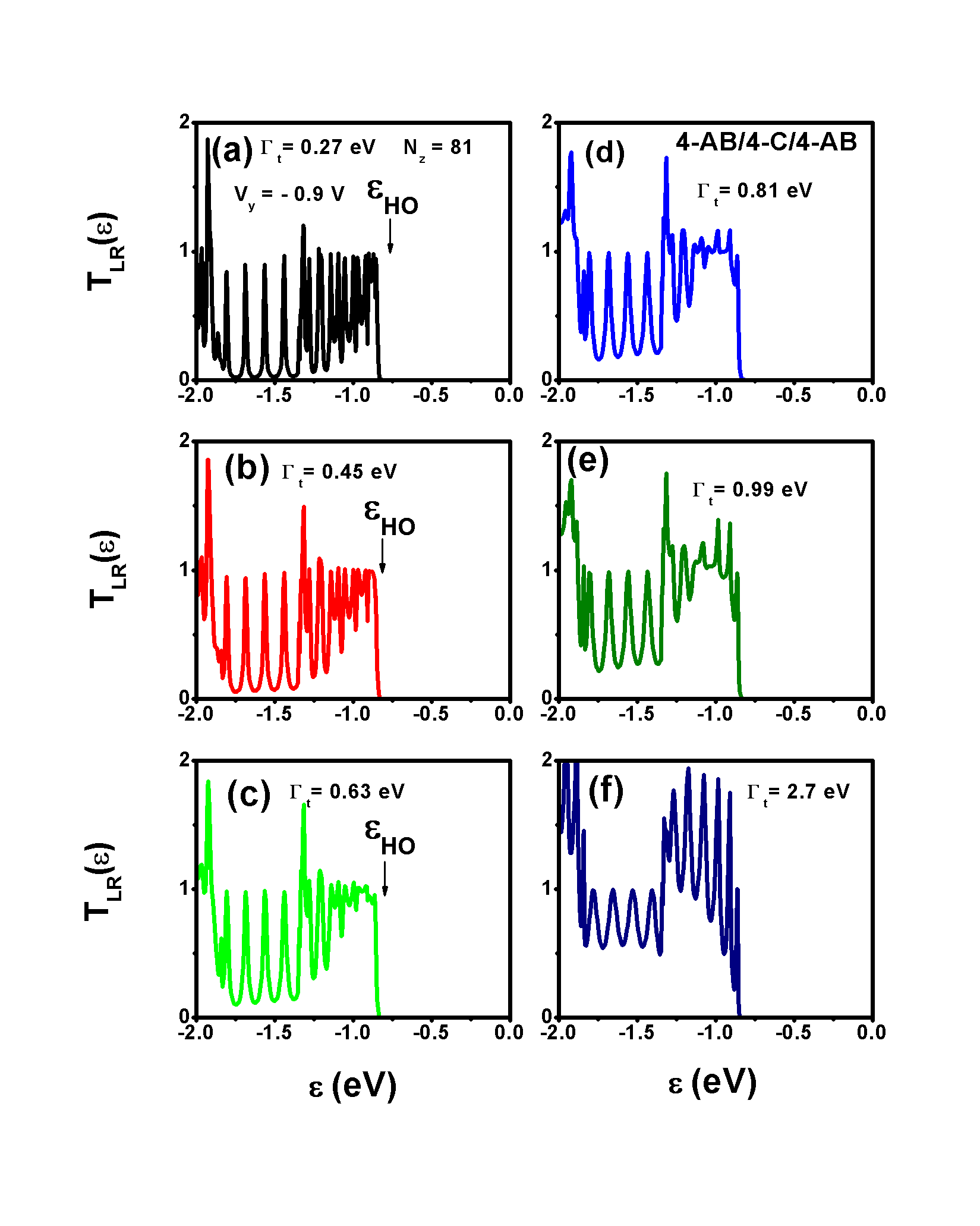}
\caption{Transmission coefficient ${\cal T}_{LR}(\varepsilon)$ of
4-AB/4-C/4-AB ZGNR segments as functions of $\varepsilon$ for
various $\Gamma_t$ values at $N_a = 12$, $N_z = 81$, $\delta =
0.9$~eV and $V_y = -0.9$~V: (a) $\Gamma_t = 0.27$~eV, (b)
$\Gamma_t = 0.45$~eV, (c) $\Gamma_t = 0.63$~eV, (d) $\Gamma_t =
0.81$~eV, (e) $\Gamma_t = 0.99$~eV, and (f) $\Gamma_t = 2.7$~eV.
Here, $N_z = 81$ corresponds to a channel length of $L_z =
9.84$~nm.}
\end{figure}

As observed in Fig. 3, the curvature of the conduction (valence)
subband in 4-AB/4-C/4-AB ZGNRs can be adjusted by a transverse
electric field. This indicates that the effective masses of
electrons (holes) and the density of states near the subband edge
depend on the applied voltage strength ($V_y$). It is expected
that the ${\cal T}_{LR}(\varepsilon)$ curves are also
significantly influenced by these transverse electric fields. To
examine the impact of $V_y$, we present the calculated ${\cal
T}_{LR}(\varepsilon)$ curves for 4-AB/4-C/4-AB ZGNR segments
across various $V_y$ values at $\Gamma_t = 2.7$~eV in Fig. 7.
According to the results of Fig. 6, the spectra of the
transmission coefficient curve at $\Gamma_t = 2.7$~eV can reach
two, illustrating the valley degeneracy. Therefore, we set
$\Gamma_t =2.7$~eV in Fig.7. The spectra of ${\cal
T}_{LR}(\varepsilon)$ depend not only on the magnitude of the
electric fields but also on their direction. Moreover, the shapes
of the ${\cal T}_{LR}(\varepsilon)$ curves vary distinctly with
each $V_y$. In reference [\onlinecite{Whitney}], it was noted that
efficient heat engines with optimized power outputs can be
realized when the ${\cal T}_{LR}(\varepsilon)$ curve exhibits an
ideal rectangular shape. However, such an ideal rectangular curve
(${\cal T}_{LR}(\varepsilon) = N_v$ for $\varepsilon <
\varepsilon_{HO}$, where $N_v$ represents valley degeneracy)
exists only in infinitely long 1D systems without defects
[\onlinecite{Areshkin}-\onlinecite{Matsuda}].

\begin{figure}[h]
\centering
\includegraphics[angle=0,scale=0.3]{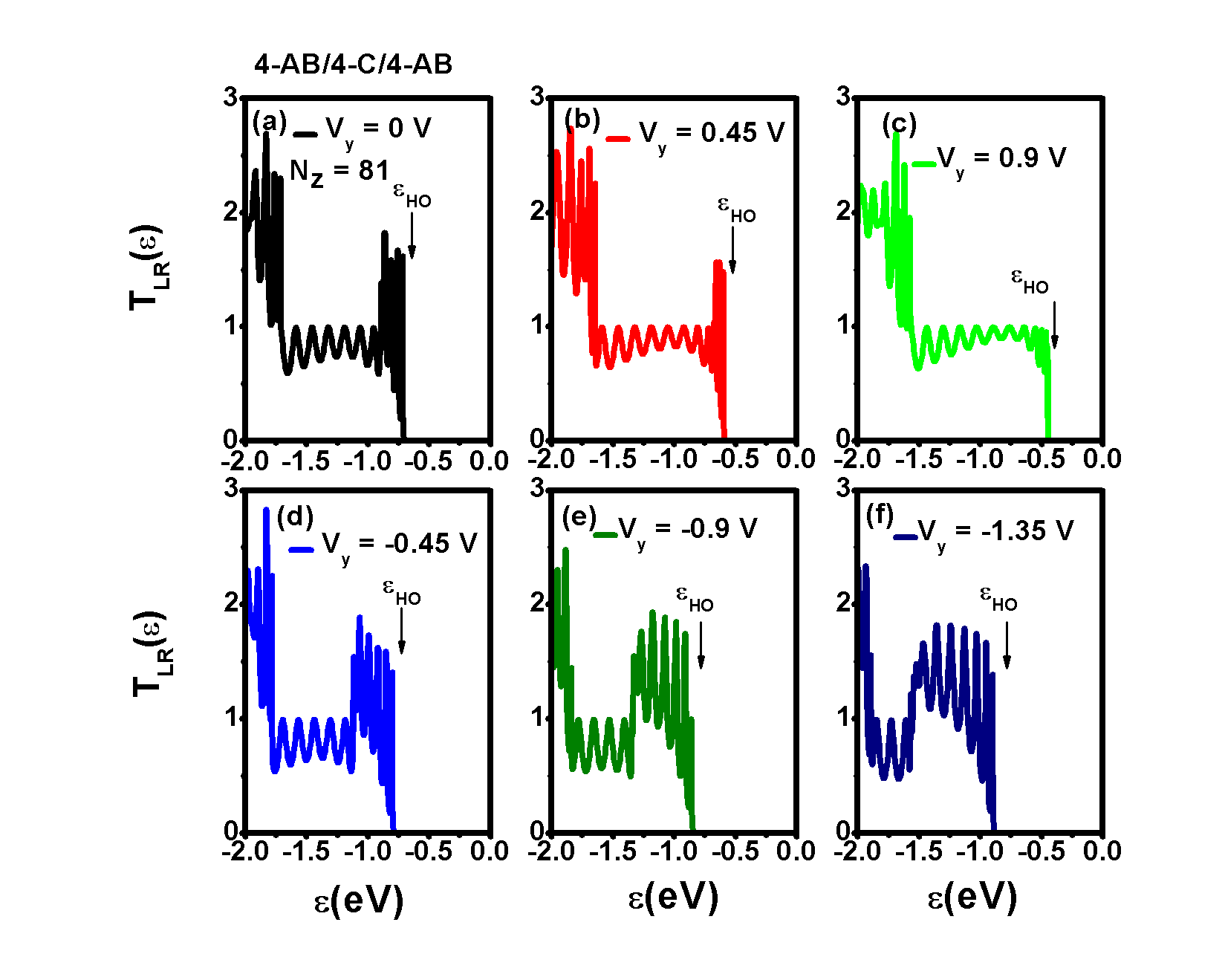}
\caption{Transmission coefficient ${\cal T}_{LR}(\varepsilon)$ of
4-AB/4-C/4-AB ZGNR segments as functions of $\varepsilon$ for
various $V_y$ values at $\Gamma_t = 2.7$~eV. Panels show (a) $V_y
= 0$~V, (b) $V_y = 0.45$~V, (c) $V_y = 0.9$~V, (d) $V_y =
-0.45$~V, (e) $V_y = -0.9$~V, and (f) $V_y = -1.35$~V. Other
physical parameters are identical to those in Fig. 6.}
\end{figure}

\subsection{Thermoelectric Properties of Finite 4-AB/4-C/4-AB ZGNR Segments}

The transmission coefficients presented in Fig. 6 and Fig. 7
reveal the electrical conductance $G_e(\mu)$ of 4-AB/4-C/4-AB ZGNR
segments at zero temperature, given by $G_e(\mu) = G_0 {\cal
T}_{LR}(\mu)$, where $G_0 = \frac{2e^2}{h}$ is the quantum
conductance. For ZGNRs with two valley degeneracy, the ideal
$G_e(\mu)$ reaches $\frac{4e^2}{h}$[\onlinecite{XieXC}]. To
illustrate the contact effects on the thermoelectric properties of
4-AB/4-C/4-AB ZGNRs, we plot the curves of electrical conductance
($G_e$), Seebeck coefficient ($S$), power factor ($PF = S^2 G_e$),
and figure of merit ($ZT$) at $T = 324$ K in Fig. 8. Throughout
this article, we use specific constants: $G_0 = 2e^2/h = 77.5~\mu
S$ for electrical conductance, $k_B/e = 86.25 \mu V/K$ for the
Seebeck coefficient, and $2k_B^2/h = 0.575pW/K^2$ for the power
factor, based on of Eq.(2).

In Fig.8(a), the curves of $G_e(\mu)$ at finite temperature are
influenced by the averaged ${\cal T}_{LR}(\varepsilon)$ curves,
with the averaging range of $\varepsilon$ determined by
$\frac{1}{4k_BT \cosh^2((\varepsilon-\mu)/(2k_BT))}$. At low
temperatures, states near $\mu$ dominate charge transport. Fig.
8(b) shows that the maximum Seebeck coefficient occurs at small
$\Gamma_t = 0.27$ eV, although the six $S$ curves nearly coincide
in the range $\varepsilon_{HO} < \mu < -0.45$ eV. In Fig. 8(c) and
8(d), the maximum power factor $PF_{max} = 1.16$ and figure of
merit $ZT_{max} = 3.46$ are observed at $\Gamma_t = 0.81$ eV. The
figure of merit $ZT$ is calculated using $ZT = \frac{S^2 G_e
T}{\kappa_e + \kappa_{ph}}$, where $\kappa_{ph}$ represents the
phonon thermal conductance of the 4-AB/4-C/4-AB structure,
approximated here as $\kappa_{ph} = F_s \kappa_{GNR}$ with
$\kappa_{GNR} = \frac{\pi^2 k_B^2 T}{3h}$. Here, $F_s = 0.1$
accounts for substrate effects, reducing $\kappa_{ph}$ as shown
theoretically for ZGNRs with boron nitride interfaces
[\onlinecite{TangXF}--\onlinecite{Algharagholy}]. Note that
$PF_{max}$ and $ZT_{max}$ for $\Gamma_t = 0.81$ eV occur at
chemical potentials $\mu = -0.823$ eV and $\mu = -0.787$ eV,
respectively. $ZT$ values exceeding three are found only within a
very small range of $\mu$. The optimized chemical potential for
maximizing the power factor is $21$ meV above $\epsilon_{HO} =
-0.844$ eV, known as thermionic-assisted transport (TAT).
Experimentally, achieving such optimized positions of electrode
chemical potentials depicted in Fig. 8 presents a challenge due to
difficulties in precise modulation[\onlinecite{Ranganayakulu}].
Introducing transverse electric fields may offer practical means
to adjust the separation between $\varepsilon_{HO}$ and $\mu$
levels.

\begin{figure}[h]
\centering
\includegraphics[angle=0,scale=0.3]{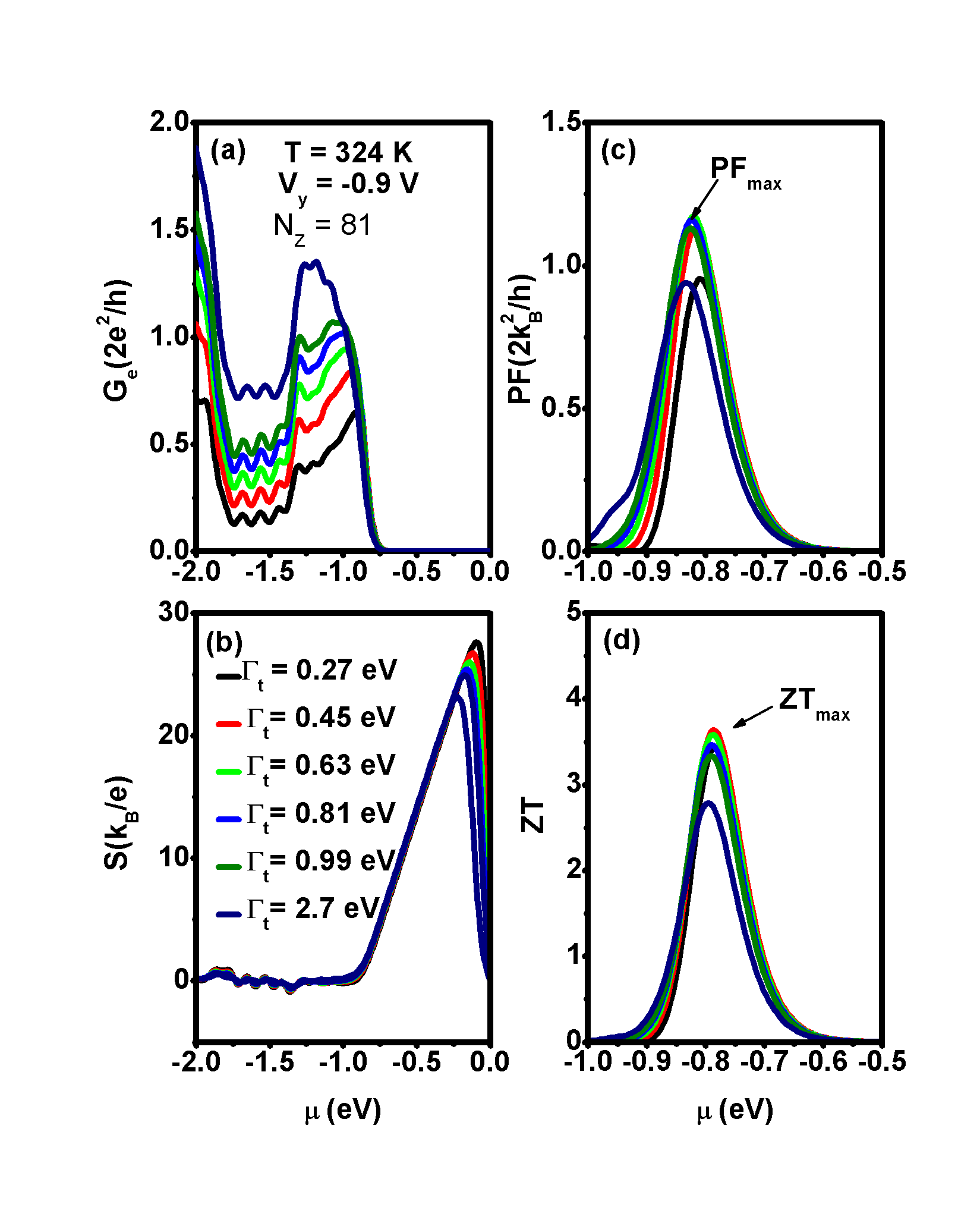}
\caption{(a) Electrical conductance $G_e$, (b) Seebeck coefficient
($S$), (c) power factor ($PF$), and (d) figure of merit ($ZT$) as
functions of chemical potential $\mu$ for different $\Gamma_t$
values at $T = 324$~K. Other physical parameters are identical to
those in Fig. 6.}
\end{figure}

According to the results in Fig. 8, the thermoelectric performance
at room temperature deteriorates when the chemical potential $\mu$
falls below $\varepsilon_{HO}$. This scenario, known as direct
ballistic transport (DBT), corresponds to $\mu <
\varepsilon_{HO}$. For an ideal scenario where ${\cal
T}_{LR}(\varepsilon) = N_v$, the Seebeck coefficients approach
zero at low temperatures when $\mu < \varepsilon_{HO}$ because $S$
does not favor the transmission coefficient curve with symmetrical
behavior around $\varepsilon \approx \mu$. This highlights the
significant impact of the shape of the ${\cal
T}_{LR}(\varepsilon)$ curve on enhancing $S$ in DBT conditions.

Therefore, we analyze the ${\cal T}_{LR}(\varepsilon)$ curves at
$V_y = -0.9$~V and $V_y = 0.9$~V shown in Fig. 7(e) and 7(c),
respectively, and present their electrical conductance, Seebeck
coefficient, and power factor as functions of temperature ($T$)
for three values of chemical potential $\Delta = \mu -
\varepsilon_{HO}$ in Fig. 9. Specifically, for $V_y = -0.9$~V, we
consider $\Delta = -20$~meV, $\Delta = -29$~meV, and $\Delta =
-38$~meV. For $V_y = 0.9$~V with $\varepsilon_{HO} = -0.4454$~eV,
we consider $\Delta = -9$~meV, $\Delta = -18$~meV, and $\Delta =
-27$~meV. In Fig. 9(a) and 9(d), the electrical conductances at
low temperatures directly reflect the transmission coefficient
spectra. As the temperature increases to and exceeds room
temperature ($k_BT > 27$~meV), the electrical conductances become
less sensitive to temperature variations.

Unlike electrical conductance, Seebeck coefficients exhibit a
temperature-dependent bipolar behavior. The positive and negative
signs reflect competition between electrons and holes,
corresponding to occupied states above $\mu$ and unoccupied states
below $\mu$, respectively. As temperature increases beyond $k_BT >
10$~meV, positive $S$ values indicate dominance of hole carriers.
In this temperature range, higher temperatures lead to larger
Seebeck coefficients, highlighting temperature-dependent
electron-hole asymmetry in DBT when $\mu$ approaches or falls
below $\varepsilon_{HO}$. For instance, the curve with $\Delta =
-20$~meV and $V_y = -0.9$V exhibits a maximum $S$ value of $138\mu
V/K$ at $k_BT = 80$meV, which is higher than the $120\mu V/K$
observed for $\Delta = -9$~meV and $V_y = 0.9$~V at the same
temperature. Consequently, the maximum power factor $PF_{max} =
1.21 \cdot \frac{2k_B^2}{h}$ at $k_BT = 80$~meV occurs for $\Delta
= -20$~meV and $V_y = -0.9$~V. To demonstrate that the results in
Fig. 9 are not significantly affected by a larger $N_z$, we
compare the transmission coefficients at $N_z = 81$ and $N_z = 101
$(with $L_a = 12.3$~nm) in Appendix B. We find that these two
curves show very little difference. This indicates that the $N_z =
81 $ used to calculate the transmission coefficients in Figs. 6
and 7 is sufficiently converged for $\Gamma_t = 2.7$~eV.

The variations in chemical potential depicted in Fig. 9 correspond
to behaviors observed in co-doped Germanium Telluride
($Ge_{1-x-y}Sb_xBi_yTe$) thermoelectric materials. It is
noteworthy that the electronic structures of $GeTe$ exhibit
multiple quasi-1D valley structures near the Fermi energy
[\onlinecite{Ranganayakulu}]. The behaviors of electrical
conductance, Seebeck coefficient, and power factor with respect to
temperature in Fig. 9(a), 9(b), and 9(c) effectively elucidate the
thermoelectric properties of these materials within the
temperature range of $10$ meV $< k_BT < 70$ meV. The arch-shaped
transmission coefficient curve provides insights into the
potential mechanism underlying the thermoelectric properties of
$GeTe$ materials with multiple valley degeneracies.

\begin{figure}[h]
\centering
\includegraphics[angle=0,scale=0.3]{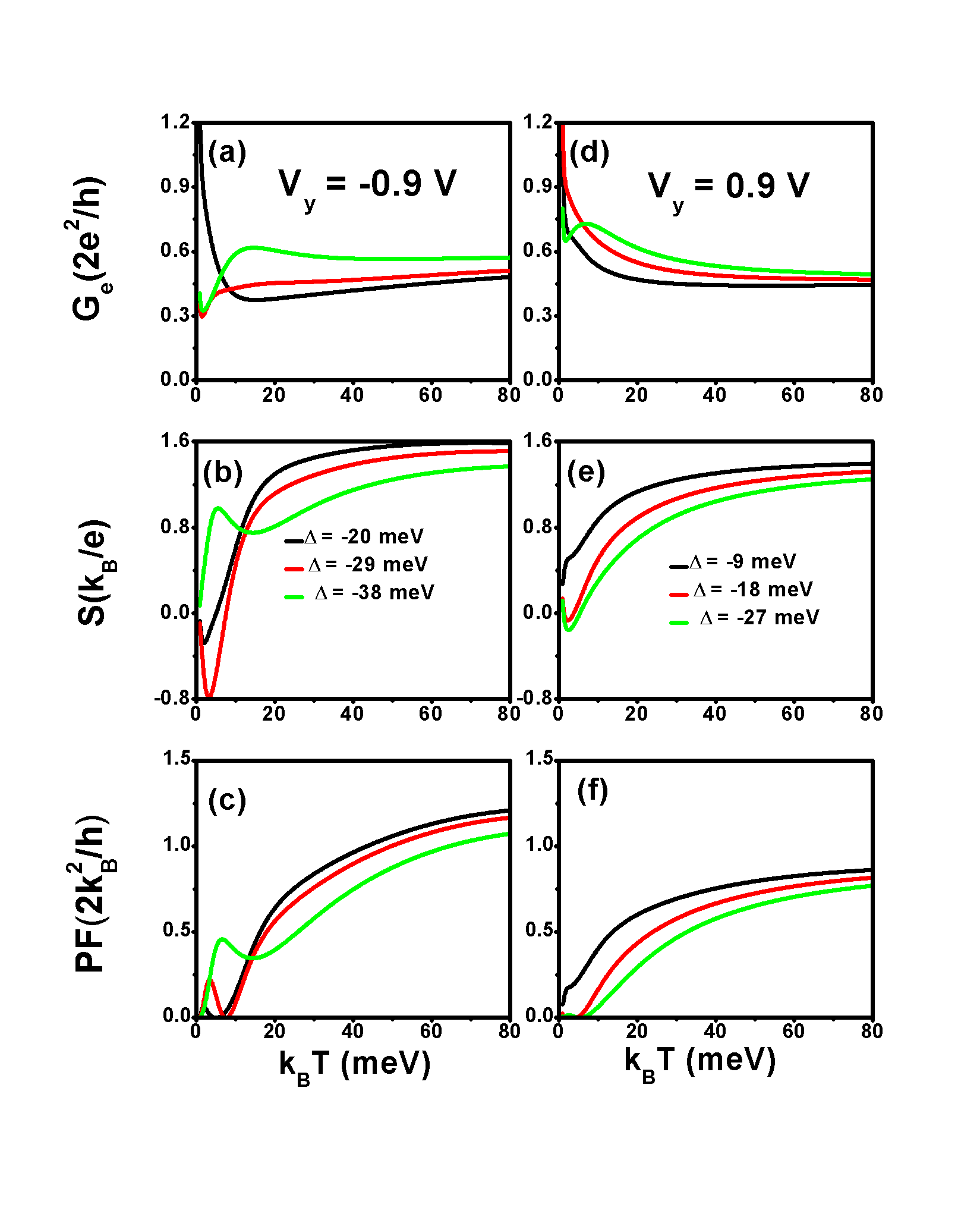}
\caption{(a) Electrical conductance ($G_e$), (b) Seebeck
coefficient ($S$), and (c) power factor ($PF = S^2 G_e$) as
functions of temperature for various chemical potential values at
$V_y = -0.9$~V. (d) Electrical conductance ($G_e$), (e) Seebeck
coefficient ($S$), and (f) power factor ($PF = S^2 G_e$) as
functions of temperature for various chemical potential values at
$V_y = 0.9$~V. The analysis is based on a 4-AB/4-C/4-AB ZGNR
segment with $N_z = 81$ ($9.84$~nm) and $\Gamma_t = 2.7$~eV.}
\end{figure}

\section{Conclusion}
We conducted theoretical studies on the electronic and
thermoelectric properties of ZGNRs with staggered sublattice
potentials described by w-AB/n-C/w-AB using the tight-binding
model and Green's function technique. The electronic structures of
w-AB/n-C/w-AB ZGNRs exhibit semiconducting phases with valley
degeneracy, resulting from band inversion of the flat bands
observed in metallic ZGNRs. This valley degeneracy is induced by
breaking inversion symmetry in the transverse direction, which can
be realized in ZGNRs stacked on separate silicon carbide or
hexagonal boron nitride sheets. The electronic structures with
valley degeneracy are further modulated by transverse electric
fields.

In 4-AB/4-C/4-AB ZGNR segments, the $|\Psi_{\ell,j}(E)|^2$ of
level degeneracy are delocalized and asymmetrically distributed
along the transverse direction. Their transmission coefficient
(${\cal T}_{LR}(\varepsilon)$) spectra are significantly
influenced by contact properties associated with electrode
materials. The shape of the ${\cal T}_{LR}(\varepsilon)$ curve can
be manipulated not only by the magnitude but also by the direction
of the applied transverse electric field, characteristics crucial
for applications in managing electron heat currents
[\onlinecite{Martinez}--\onlinecite{Kuo5}].

The thermoelectric efficiency of 4-AB/4-C/4-AB ZGNRs with
optimized electrical power output favors thermionic-assisted
transport (TAT). Level degeneracy enhances electrical conductance
while maintaining Seebeck coefficients, thereby enhancing the
power factor. In TAT, $ZT$ can also be improved due to the
enhanced power factor under conditions where $\kappa_{ph} \gg
\kappa_e$. Conversely, in direct ballistic transport (DBT), the
power factor is reduced due to the small Seebeck coefficient at
low temperatures. As $\mu$ approaches and falls below
$\varepsilon_{HO}$, significant enhancements in the Seebeck
coefficient at high temperatures and modest reductions in
electrical conductance lead to enhanced power factors. This
phenomenon results from increased electron-hole asymmetrical range
with rising temperature in 4-AB/4-C/4-AB ZGNRs. The results in
Fig. 9(a), 9(b), and 9(c) within the range of $k_BT = 20$~meV to
$k_BT = 70$~meV provide valuable insights into the thermoelectric
properties observed in materials like Germanium Telluride
[\onlinecite{Ranganayakulu}].


{}

{\bf Acknowledgments}\\
This work was supported by the National Science and Technology
Council, Taiwan under Contract No. MOST 107-2112-M-008-023MY2. We
acknowledge the use of the computing facility provided by the
Research Center for Applied Sciences, Academia Sinica, Taiwan.

\mbox{}\\
E-mail address: mtkuo@ee.ncu.edu.tw\\

 \numberwithin{figure}{section} \numberwithin{equation}{section}

\setcounter{section}{0}
\setcounter{equation}{0} 

\mbox{}\\

\appendix
\numberwithin{figure}{section}

\section{Electronic Band Structures of ZGNRs with w-AB/n-C/w-AB
Scenario}

To further understand the electronic properties depicted in Fig.
2(a), we present the electronic structures of 4-AB/4-C/4-AB for
various $\delta$ values in Fig. A.1. In Fig. A.1(a) with $\delta =
0$, the electronic structure of 4-AB/4-C/4-AB exhibits metallic
properties of ZGNRs, characterized by flat bands near the CNP
[\onlinecite{Nakada}-\onlinecite{Wakabayashi}]. In Fig. A.1(b),
with $\delta = 0.45$~eV, only the frontier conduction and valence
subbands are perturbed. Other subbands remain largely unaffected,
and a small gap separates the conduction subband from the valence
subband due to the presence of staggered sublattice potential.
Additionally, the originally flat band observed at $\delta = 0$ is
slightly distorted.

Fig. A.1(c) shows the scenario with $\delta = 0.9$~eV, where the
frontier conduction and valence subbands exhibit clear valley
structures, albeit shallow ones. Tuning $\delta$ further from
$\delta = 1.35$~eV to $\delta = 2.25$~eV leads to a significant
band inversion in the frontier conduction and valence subbands of
ZGNRs. This band inversion mechanism shown in Fig. A.1 does not
involve spin-orbit coupling, which is a mechanism often cited for
band inversion in materials such as $BiTe$ thermoelectrics.

\begin{figure}[h]
\centering
\includegraphics[angle=0,scale=0.3]{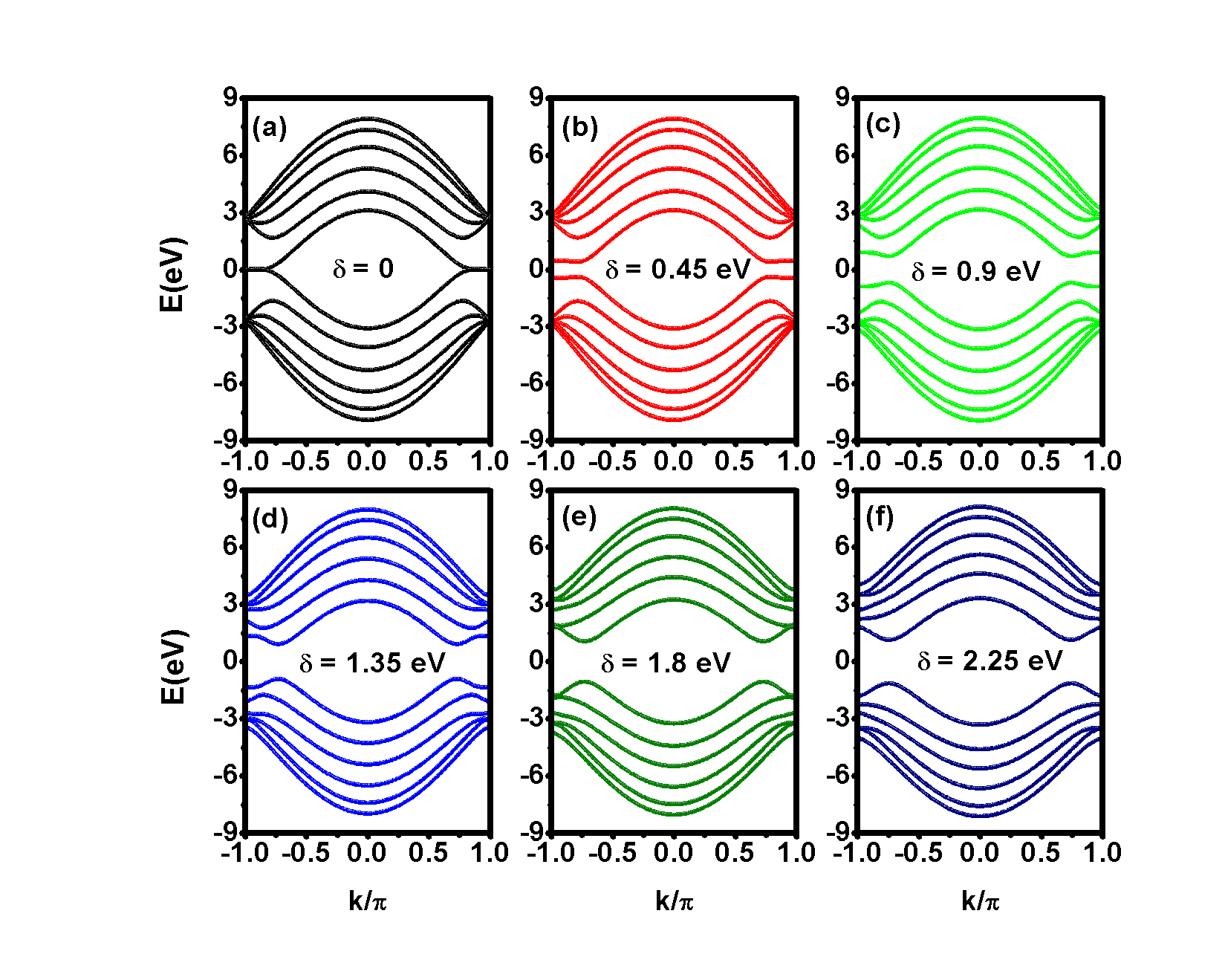}
\caption{Electronic structures of 4-AB/4-C/4-AB ZGNRs for various
$\delta$ values.}
\end{figure}

Next, we present the calculated electronic structures of
4-AB/n-C/4-AB with a fixed $\delta = 0.9$~eV for different values
of $n$ in Fig. A.2. As the value of $n$ increases, the band gaps
of ZGNRs decrease. While deeper valley structures are enhanced
with increasing $n$, the smaller band gap tends to suppress the
Seebeck coefficient at high temperatures. Additionally, wider
ZGNRs can increase phonon thermal conductance. This study focuses
specifically on the case of 4-AB/4-C/4-AB ZGNRs.

\begin{figure}[h]
\centering
\includegraphics[angle=0,scale=0.3]{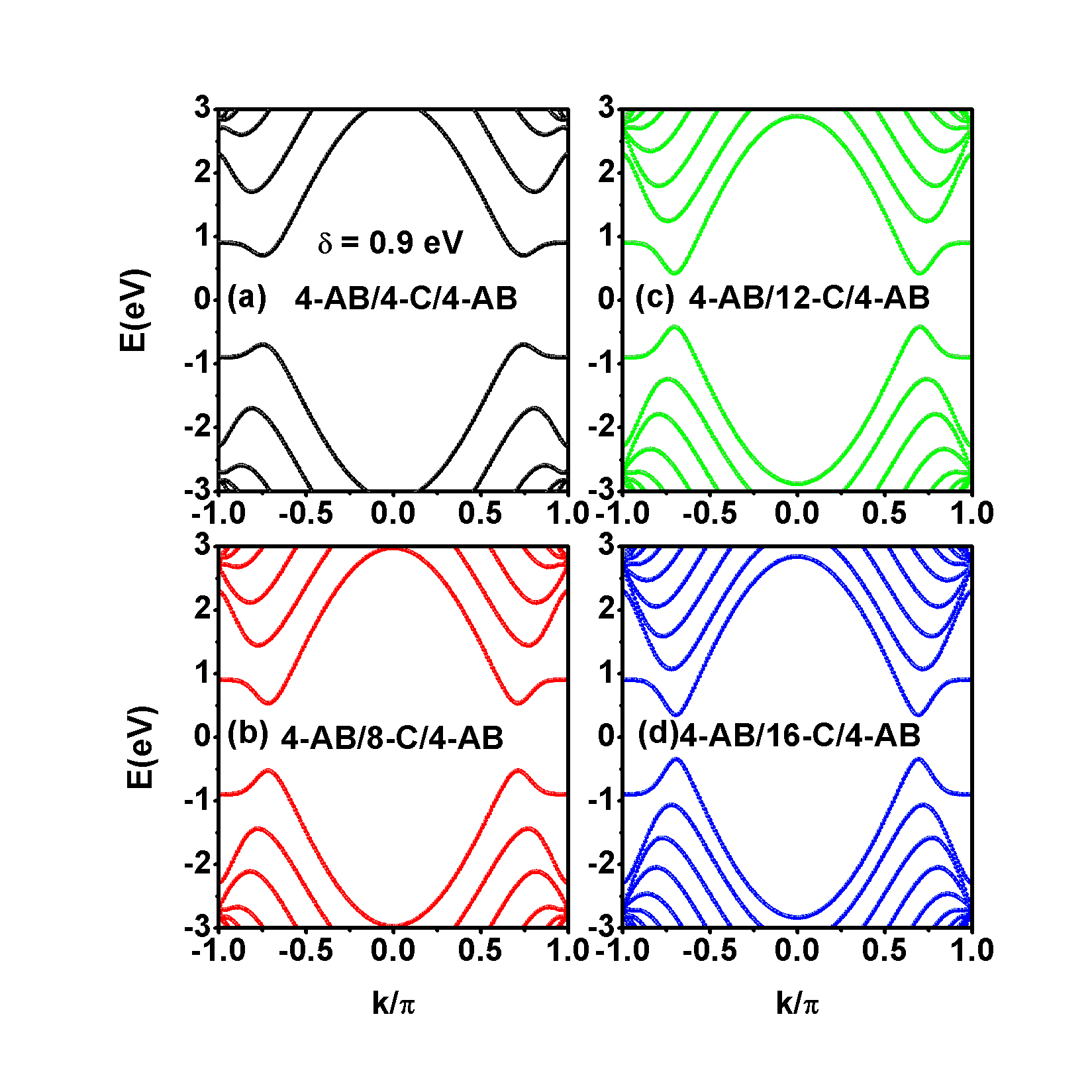}
\caption{Electronic structures of 4-AB/n-C/4-AB ZGNRs with $\delta
= 0.9$~eV for various n values.}
\end{figure}

\section{Transmission coefficients of 4-AB/4-C/4-AB ZGNRs under
transverse electric fields} \numberwithin{equation}{section}

The thermoelectric quantities in Fig. 8 and Fig. 9 are calculated
using $N_z = 81$. We aim to demonstrate that the results in these
figures do not depend on the length $L_z$. In Fig. B.1, we present
the transmission coefficient for 4-AB/4-C/4-AB ZGNRs with $N_z =
101$ (where $L_z = 12.3$~nm). The transmission coefficient curve
exhibits electron-hole symmetry, and the solid curve ($N_z = 101$)
is difficult to distinguish from the dashed curve ($N_z = 81$).
This implies that the length of the 4-AB/4-C/4-AB ZGNR segment
with $N_z = 81$ is sufficiently long at $\Gamma_t = 2.7$~eV.

\begin{figure}[h]
\centering
\includegraphics[angle=0,scale=0.3]{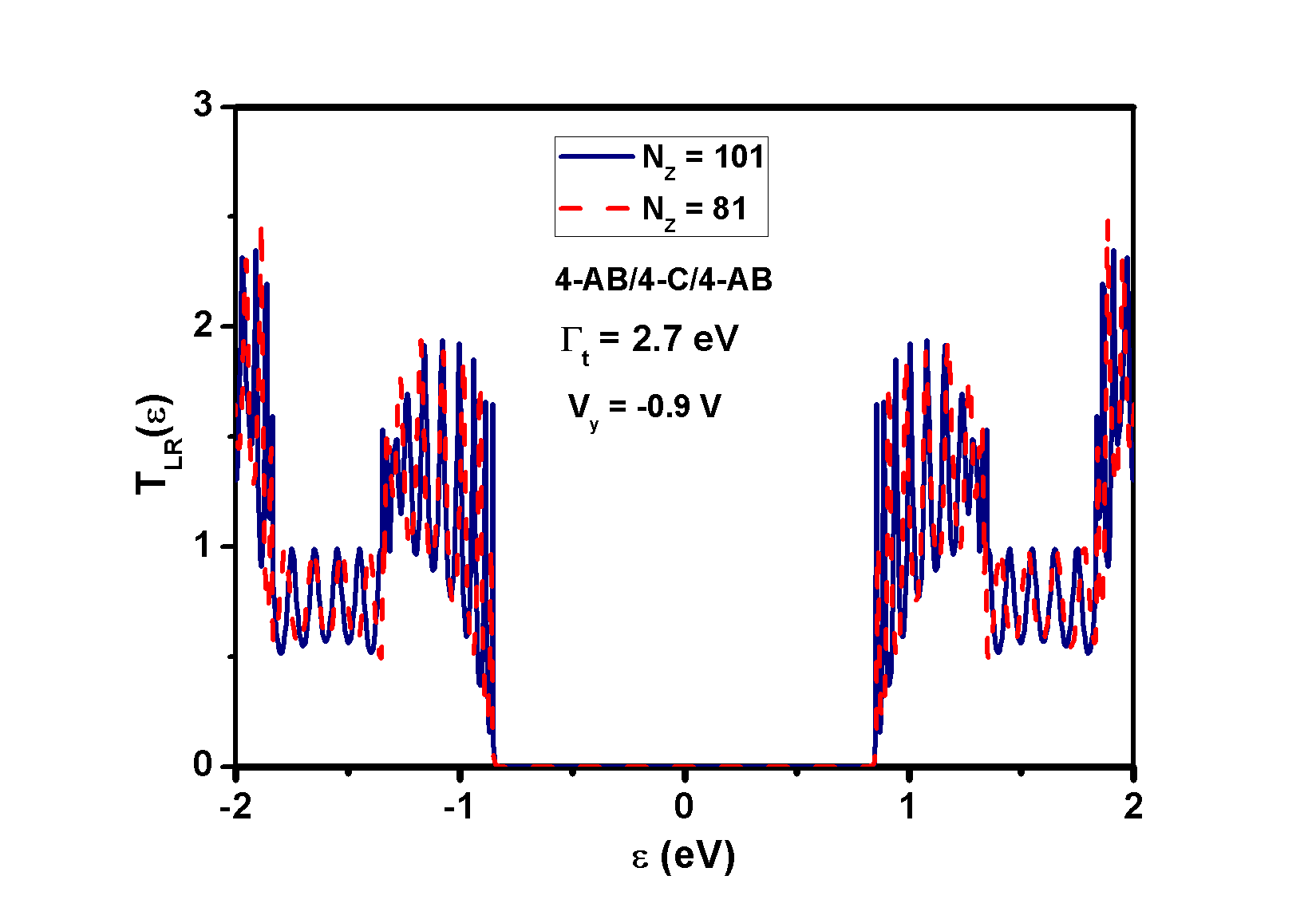}
\caption{Transmission coefficient of 4-AB/4-C/4-AB ZGNRs with
$\delta = 0.9$~eV, $V_y = -0.9$~V, and $\Gamma_t = 2.7$~eV.}
\end{figure}



\end{document}